# Minority-Spin Impurity Band in n-Type (In,Fe)As: A Materials Perspective for Ferromagnetic Semiconductors

Masaki Kobayashi[1,2,3,4], Le Duc Anh[4,5,6], Jan Minár[7], Walayat Khan[8], Stephan Borek[9], Pham Nam Hai[3,4,10], Yoshihisa Harada[11], Thorsten Schmitt[1], Masaharu Oshima[2], Atsushi Fujimori[12,13], Masaaki Tanaka[3,4], and Vladimir N. Strocov[1,]

[1]*Swiss Light Source, Paul Scherrer Institut, CH-5232 Villigen PSI, Switzerland*
[2]*Department of Applied Chemistry, School of Engineering, The University of Tokyo, 7-3-1 Hongo, Bunkyo-ku, Tokyo 113-8656, Japan*
[3]*Department of Electrical Engineering and Information Systems, The University of Tokyo, 7-3-1 Hongo, Bunkyo-ku, Tokyo 113-8656, Japan*
[4]*Center for Spintronics Research Network, The University of Tokyo, 7-3-1 Hongo, Bunkyo-ku, Tokyo 113-8656, Japan*
[5]*Institute of Engineering Innovation, Graduate School of Engineering, The University of Tokyo, 7-3-1 Hongo, Bunkyo-ku, Tokyo 113-8656, Japan*
[6]*PRESTO, JST, 4-1-8 Honcho, Kawaguchi, Saitama, 332-0012, Japan*
[7]*NewMaterials -Technologies: Research Centere, University of West Bohemia, Pilsen, Czech Rep.*
[8]*Bacha Khan University, Charsadda, KPK, Pakistan*
[9]*Deutsches Zentrum für Luft- und Raumfahrt (DLR), Oberpfaffenhofen, 82234 Weßling, Germany*
[10]*Department of Physical Electronics, Tokyo Institute of Technology, 2-12-1 Ookayama, Meguro-ku, Tokyo 152-0033, Japan*
[11]*Insititute for Solid State Physics, The University of Tokyo, 1-1-1 Koto, Sayo, Hyogo 679-5198, Japan*
[12]*Department of Physics, The University of Tokyo, 7-3-1 Hongo, Bunkyo-ku, Tokyo 113-0033, Japan*
[13]*Department of Applied Physics, Waseda University, Okubo, Shinjuku, Tokyo 169-8555, Japan*

**Abstract**

Fully understanding the properties of *n-type* ferromagnetic semiconductors (FMSs), complementary to the mainstream p-type ones, is a challenging goal in semiconductor spintronics because ferromagnetism in n-type FMSs is theoretically non-trivial. Soft-x-ray angle-resolved photoemission spectroscopy (SX-ARPES) is a powerful approach to examine the mechanism of carrier-induced ferromagnetism in FMSs. Here our SX-ARPES study on the prototypical n-type FMS (In,Fe)As reveals the entire band structure including the Fe-3*d* impurity bands (IBs) and the host InAs ones, and provides direct evidence for electron occupation of the InAs-derived conduction band (CB). A minority-spin Fe-3*d* IB is found to be located just below the conduction-band minimum (CBM). The IB is formed by the hybridization of the unoccupied Fe-3*d* states with the occupied CBM of InAs in a spin-dependent way, resulting in the large spin polarization of CB. The band structure with the IB is varied with band filling, which cannot be explained by the rigid-band picture, suggesting a unified picture for realization of carrier-induced ferromagnetism in FMS materials.

**Introduction**

Evolution of information technology has been driven by high-performance electronic devices based on high-quality semiconductor materials. Since the sophistication of semiconductor electronic devices by miniaturization has approached the technical limit, the development of new functional spin-based devices has been pursued [1, 2]. Spintronics is a research field aiming at manipulating and utilizing both the charge and spin degrees of freedom of carriers, and spintronic devices have a potential to dramatically reduce the power consumption and realize new functionalities related to the spin degree of freedom [1–5]. For instance, bilayer thin films consisting of a ferromagnetic layer and superconductors, topological insulators, or two-dimensional electron gas have recently attracted much attention for their novel spintronics functionalities, including the realization of Majorana Fermions for quantum computing [3]. To apply these spintronics functionalities to the well-established semiconductor technology, materials having both the semiconducting and ferromagnetic properties are highly desirable. Ferromagnetic semiconductors (FMSs), in which the cation sites in a semiconductor crystal are partially replaced by magnetic atoms, bear a high promise for their applications in semiconductor spintronics because of their capability to manipulate both the charge and spin degrees of freedom of carriers [4, 5]. Thus, FMSs are key materials to archive the practical application of the spin-related functionalities [1, 4, 5, 6, 7, 8].

Fundamental understanding of the origin or mechanism of ferromagnetism in FMSs is important for the application of the FMS materials to spintronics devices. It is known that the ferromagnetism in many FMSs is induced by doping carriers. In contrast to the traditional Mn-doped FMSs such as (In,Mn)As and (Ga,Mn)As that exhibit only p-type conduction, novel Fe-doped III-V FMSs have recently attracted much attention, because Fe-doped FMSs can accommodate both n- and p-type carriers and exhibit ferromagnetism with high Curie temperature $T_C$ (> 300 K) [5, 9, 10]. Using the p-type and/or n-type Fe-doped FMSs, spintronics devices have already been demonstrated [6, 7, 11, 12]. Fe-doped FMS $(In_{1-x},Fe_x)$As co-doped with Be (referred to as (In,Fe)As:Be hereafter) is the first n-type FMS [13, 14], where the Be ions act as double donors [15]. (In,Fe)As:Be shows ferromagnetic properties when its electron-carrier concentration is higher than the threshold of electron concentration $6 \times 10^{18}$ cm$^{-3}$ [13]. The findings of light electron effective mass [14] and large spin splitting in the conduction-band minimum (CBM) $\Delta_{ex}$ [12] suggest that highly mobile conduction electrons are spin-polarized in the conduction band (CB) of (In,Fe)As:Be. Although the carrier-induced ferromagnetic properties and band structure of (In,Fe)As have been intensively studied so far, the precise mechanism of the ferromagnetism still remains an enigma.

Knowledge of the electronic structure, which is composed of the band structure of the host semiconductor and the 3$d$ impurity band (IB), is indispensable for understanding the mechanism of the ferromagnetism in FMSs. Theoretically, it was considered impossible or difficult to realize a zinc blend n-type FMS due to the weak orbital mixing between the host $s$ and impurity $d$ orbitals [16]. In this Letter, we have investigated the entire band structure of the prototypical n-type FMS (In,Fe)As:Be using SX-ARPES and spin-density-functional-theory (SDFT) calculation to understand the origin of the carrier-induced ferromagnetism. SX-ARPES is one of the most powerful experimental techniques to study the electronic structures of FMSs [17, 18, 19]. By using SX-ARPES, we revel the entire band structure of (In,Fe)As:Be including the Fe-3$d$ IB and electron occupation of the InAs-derived CB. Based on the experimental findings, we have identified that the Fe-3$d$ IB located near CBM is formed by the hybridization of the CB of InAs and the Fe 3$d$ state in a spin-dependent way, resulting in the large $s$-$d$ exchange splitting in (In,Fe)As:Be. The underlying physics of the formation of the IB states in n-type (In,Fe)As is found to be similar to that in p-type (Ga,Mn)As, providing a unified picture of ferromagnetism in FMSs irrespective to the carrier types. This suggests a materials perspective to realize the carrier-induced ferromagnetism in FMSs.

**Experimental**

In$_{1-x}$Fe$_x$As:Be ($x$ = 0.05), In$_{1-x}$Fe$_x$As ($x$ = 0.05), and InAs:Be thin films with a thickness of 20 nm were grown on InAs(001) substrates at 240 °C in an ultra-high vacuum by molecular beam epitaxy (MBE). The Be concentration was ~2×10$^{19}$ cm$^{-3}$. In order to avoid surface oxidation, the thin films were covered by an amorphous As passivation layer with a thickness of ~1 nm after the MBE growth. The SX-ARPES measurements were performed directly through the As capping layer due to the increase of probing depth in the SX region. This has allowed us to avoid heating of the sample to evaporate the As layer and therefore avoid risks of FeAs second-phase precipitations. This experimental methodology has already been established in our previous works on (Ga,Mn)As [20]. The magnetic properties of the samples were confirmed by the Arrott plot of the magnetic circular dichroism intensity and the $T_C$ of the (In$_{0.95}$,Fe$_{0.05}$)As:Be film was estimated to be

~40 K. The SX-ARPES experiments were performed at the ADRESS beamline of SLS, which delivers SX radiation with variable polarization in the energy range from 300 to 1600 eV. A high photon flux above $10^{13}$ photons/s/0.01%BW at 1 keV delivered by the beamline combine with grazing-incidence experimental geometry of the SX-ARPES end station [21] allowed us to compensate a dramatic loss of the valence-band cross section in the SX energy range, in our case particularly severe in view of small atomic concentrations of Fe impurities. The sample was cooled to 10.7 K by liquid He to suppress the reduction of the coherent ***k***-resolved spectral component at high energies. The measurements were performed mostly at energy resolution $\Delta E \sim 120$ meV. In our experimental geometry [21], the parity of the photo-excitation operator, $\boldsymbol{A} \cdot \boldsymbol{p}$, with respect to the beam-detector plane was odd for the horizontal and even for the vertical polarizations. Our navigation in ***k***-space considering the photon momentum $k^{\text{ph}} = h\nu/c$ as described in Ref. [21] in relation to our experimental geometry. Details of the SDFT calculations are described in Supplementary Material [22].

**SX-ARPES**

ARPES in the SX photon energy range enables us to reveal the bulk band structures of single-crystal materials with sharp definition of three-dimensional electron wavevector ***k*** [20]. Figure 1(a) shows the out-of-plane Fermi surface mapping in the $k_x$-$k_z$ plane of an (In$_{0.95}$Fe$_{0.05}$)As:Be thin film capped with an amorphous As layer. Here, the $k_z$ and $k_x$ directions are normal and parallel to the in-plane Γ-K-X symmetry line of the Brillouin zone, respectively. The mapping shows tiny but clear Fermi surface around the Γ points. As shown in Fig. 1(a), clear $k_z$ dispersion of all the surface contours excludes the presence of any surface states that are identified by the absence of their $k_z$ direction. This ensures that our SX-ARPES spectra truly reflect the bulk electronic structure of (In,Fe)As:Be. Figure 1(b) shows an enlarged plot of the Fermi surface around the Γ point [$(k_x, k_z)$ = (-1.0, 15) $(2\sqrt{2}\pi/a)$]. The cross-sections of the Fermi surface in the $k_x$-$k_z$ out-of-plane as well as that in the $k_x$-$k_y$ in-plane (not shown) are circular. Actually, the Fermi momenta are about 0.12, 0.12, and 0.13 Å$^{-1}$ for the $k_x$, $k_y$ (parallel to the Γ-X line), and $k_z$ directions, respectively. This suggests that the shape of the Fermi surface is spherical within the experimental accuracy in (In,Fe)As:Be. The electron carrier concentration estimated from the volume of the Fermi surface is ~2×10$^{19}$ cm$^{-3}$; this value is above the threshold of electron concentration (~6×10$^{18}$ cm$^{-3}$) to induce the ferromagnetism in (In,Fe)As [13], and it is in good agreement with the carrier concentration estimated from the Hall measurement and is of the same order of magnitude as the concentration of co-doped Be. This result is consistent with the assumption that the Be dopant acts as a double donor in (In,Fe)As. Figure 1(c) shows the band dispersion along the Γ-K-X symmetry line. Here, the incident photon energy $h\nu$ is set at 908 eV in order to bring the $k_z$ value to the Γ point (see Fig. 1(a)). The hole-like band dispersions centered at the Γ point are the light-hole (LH) and split-off (SO) bands (see Supplemental Material I for the linear polarization dependence [22]), and the electron-like one near the Fermi level ($E_F$) centered at Γ identifies the CB. As shown in Figs. 1(c) and 1(d), the CB dispersion clearly crosses $E_F$ and forms the small Fermi surface around the Γ point so-called electron pocket in (In,Fe)As:Be, consistent with the n-type conductivity confirmed by Hall effect and thermoelectric Seebeck effect measurements [13]. These results demonstrate that the electronic transport in the ferromagnetic (In,Fe)As:Be arises from the electron carriers in the CB originating from the $s$ band of InAs. This naturally explains their light effective mass [14] and the long mean-free path of electron carriers [11].

**Resonant ARPES**

In addition to the band dispersions, the location of the Fe-3$d$ IB is also important to understand the origin of the ferromagnetism in (In,Fe)As. To identify the Fe-3$d$ IB, we have employed resonant ARPES (rARPES) at the Fe $L_3$ edge. The resonant enhancement of the photoemission intensity occurs at the incident photon energy $h\nu$ corresponding to the peaks of the Fe $L_3$ x-ray absorption spectroscopy (XAS) spectrum (Fig. 2(a)). Figure 2(b) shows the on- and off-resonance ARPES spectra measured with $h\nu$ = 708.2 eV (the red arrow in Fig. 2(a)) and 705 eV (the black arrow), respectively. Note that, in addition to the intense Fe-3$d$ IB ($\beta$-IB), a weak Fe-3$d$ IB ($\alpha$-IB) appears near the CBM. The intensity of the $\alpha$-IB is resonantly enhanced when $h\nu$ is at the peak of the XAS spectrum (see Fig. 2(a)), indicating that the $\alpha$-IB is derived from the Fe 3$d$ orbitals in (In,Fe)As, not from extrinsic Fe oxides. The $\beta$-IB results from the hybridization of the Fe 3$d$ orbitals with the valence-band (VB) and appears in the vicinity of the valence-band maximum (VBM). While the Mn-3$d$ IB in (Ga,Mn)As splits off from the VBM due to the strong $p$-$d$ hybridization [17], the Fe-3$d$ IB ($\beta$-IB) in (In,Fe)As does not. To elucidate the hybridization between the Fe 3$d$ orbital and the ligand experimentally, the Fe-3$d$ partial density of states (PDOS) corresponding to the difference between the on- and off-resonance spectra are plotted in Fig. 2(c). In addition to the $\boldsymbol{k}$-independent $\alpha$-IB and $\beta$-IB (marked by blue and pink translucent vertical belts), the PDOS around $k_x$ = −1.0 $(2\sqrt{2}\pi/a)$ shows a dispersive feature corresponding to the dispersion of the LH band (marked by vertical bars). Since the Fe 3$d$ orbitals are independent of $\boldsymbol{k}$, the observation of dispersion in the PDOS evidences that the LH band is also resonantly enhanced at the Fe $L_3$ absorption edge through hybridization with the Fe 3$d$ orbital.

**Comparison between ferromagnetic and paramagnetic (In,Fe)As**

The band structure of (In,Fe)As:Be revealed by SX-ARPES is summarized in Fig. 3(a), where the band dispersions along the Γ-K-X line are superimposed with the Fe-3$d$ IBs. To separate the effects of Fe doping on the band structure of (In,Fe)As from that of Be, we have measured SX-ARPES spectra of the host compound InAs:Be grown under the same condition, as presented in Fig. 3(b). The small Fermi surface around the Γ point also exists in Be-doped InAs. Comparing the experimental band dispersions between InAs:Be and (In,Fe)As:Be, we find them nearly identical except for the presence of IBs in (In,Fe)As:Be. These results suggest that the electron carriers in (In,Fe)As:Be originate predominantly from the Be co-doping and not from the Fe doping. This indicates that the Fe ions substitute for the In cation sites in the $Fe^{3+}$ valence and provide the system with only the magnetic moments. To reveal the carrier concentration dependence of the band structure, an $(In_{0.95},Fe_{0.05})$As film without Be doping have been measured. Figure 3(c) shows the band structure measured on the paramagnetic $(In_{0.95},Fe_{0.05})$As film. One can see that both the electron pocket and the $\alpha$-IB disappear in the (In,Fe)As sample with a lower electron carrier concentration. If the rigid-band model is applicable, because the position of $E_F$ is at or slightly below the CBM in the paramagnetic (In,Fe)As, the $\alpha$-IB located just below the CBM would be occupied, but it is not the case (see Fig. 3(c)). The present observation suggests that the location of the $\alpha$-IB depends on the electron carrier concentration.

Based on these experimental findings, schematic views of the density of states (DOS) of ferromagnetic (In,Fe)As:Be and paramagnetic (In,Fe)As are shown in Figs. 4(a) and 4(b), respectively. It should be mentioned here that the concentration of the Be dopant is one to two orders of magnitude smaller than that of the Fe ions. If an in-gap Fe-3$d$ IB exists below the CBM, the amount of the supplied electrons by the Be doping was insufficient to fully occupy the IB and $E_F$ would be located in the IB, not in the CB. Considering appearance of the $\alpha$-IB depending on the carrier concentration, the electron density of the $\alpha$-IB located below $E_F$ is presumably the same order of the Be dopant of $10^{19}$ cm$^{-3}$, suggesting the partial occupation of the $\alpha$-IB as show in Fig. 4(a). This phenomenon looks analogous to the evolution of an IB state located below $E_F$ with increasing K$^+$ concentration in K$_x$C$_{60}$ [23, 24]. A fundamental question here is why the partially filled $\alpha$-IB is located below the CBM in (In,Fe)As:Be despite the electron-carrier occupation of the CBM.

**SDFT calculation**

As shown in Fig. 2(a) the Fe $L_3$ XAS is characterized by a smooth and broad spectral peak without multiplet structure, which is similar to that of Fe metal or metallic Fe compounds (see also Supplemental Material II [22]). Because the Fe 3$d$ states in (In,Fe)As:Be are likely to be strongly covalent with ligand As $p$ orbitals and are different from purely ionic Fe$^{3+}$ [25], the broad spectral feature of the XAS spectrum reflects the strong hybridization of the Fe 3$d$ states with the ligand. As shown in Fig. S3 (see Supplemental Material III [22]), the SDFT calculations have demonstrated that the Fe 3$d$ minority-spin (↓) states strongly hybridize with the ligand VB and CB states. Only the minority-spin states of the ligands hybridize with the Fe 3$d$ states, because the majority-spin (↑) 3$d$ states are fully occupied and located well below the host valence band. The $\beta$-IB originates from the hybridization between the Fe 3$d_\downarrow$ states and the VBM states. Note that the calculated $e_\downarrow$ states show a slightly dispersive feature, indicating finite hybridization of the $e_\downarrow$ states with the minority-spin host band states. This result indicates that the hybridization between the $e_\downarrow$ and CBM states forms the $\alpha$-IB just below the CBM as shown in Fig. 4(d), if the hybridization is strong enough.

**Discussion: Origin of the ferromagnetism in n-type (In,Fe)As**

With the increase of the carrier concentration $n$, the $\alpha$-IB are firstly occupied partially and then the carriers also subsequently occupy the CBM because the lowest energy of the $\alpha$-IB should be located below the CBM. When the CB is partially occupied, it is probable that the $\alpha$-IB splits off from the CBM through the $s$-$e_\downarrow$ hybridization, leading to the energy gap between the occupied and unoccupied $\alpha$-IB states, as shown in Fig. 4(c). This partially filled split-off IB below $E_F$ is similar to the formation of IB in the band gap observed in $K_xC_{60}$, where the IB is derived from electrons donated into the lowest unoccupied molecular orbital by the potassium [23, 24]. Additionally, emergence of the split-off IB related to the carrier-induced ferromagnetism is common feature to p-type FMS (Ga,Mn)As [17]. It should note here that the $s$-$e_\downarrow$ hybridization is spin-dependent in (In,Fe)As, that is, it happens only for the minority spin (↓) states due to the majority spin (↑) states of Fe $3d$ are fully occupied. With the partially filled split-off $\alpha$-IB, the minority-spin (↓) CBM loses states whereas the majority-spin (↑) one does not, as shown in Fig. 4(c).

The formation of the split-off state in the spin-dependent way results in a large spin-splitting $\Delta_{ex}$ of the CBM in (In,Fe)As, as shown in Fig. 4(c). Indeed, the spin-splitting of the CB as large as 30 – 50 meV has been observed in the n-(In,Fe)As/p-InAs Esaki-diode structures [12]. According to the Anderson Hamiltonian [26], the value of the $s$-$d$ exchange interaction $N_0\alpha$ is given by $|N_0\alpha| = -2|V_{sd}|^2 \left(\frac{1}{E_C-\varepsilon_d} + \frac{1}{U-E_C+\varepsilon_d}\right)$, where $E_C$ is the energy of the CBM, $\varepsilon_d$ is the energy of the $d$ states, $U$ is the Coulomb repulsion between electrons in the $d$ level of Fe, and $V_{sd}$ is the $s$-$d$ mixing potential [27]. From the value of $T_C$, $|N_0\alpha|$ for (In,Fe)As has been estimated to be ~ 2.8 eV [14], which is much larger than the $p$-$d$ exchange interaction $|N_0\beta|$~1.2 eV in (Ga,Mn)As [28]. Since $|N_0\alpha|$ is inversely proportional to the energy difference $|E_C - \varepsilon_d|$ or $|U - E_C + \varepsilon_d|$, if one of the energy differences approaches zero, the value of $|N_0\alpha|$ will be significantly enhanced. This condition is referred to as the "resonance" condition [14]. As discussed above, while $\varepsilon_d$ is well above $E_C$, the energy difference $|U - E_C + \varepsilon_d|$ is possibly small to satisfy the resonance condition in (In,Fe)As:Be. It follows from the above argument that the strong $s$-$d$ exchange interaction $|N_0\alpha|$ [12, 14] leading to the carrier-induced ferromagnetism in (In,Fe)As:Be originates from the hybridization between the minority-spin $e_\downarrow$ state and CBM states.

**A Materials Perspective for Ferromagnetic Semiconductors**

The present results put forward important aspects for the materials design of FMSs. In general, the ionicity of doped transition-metal (TM) ions will increase with the band gap of the host semiconductor [29]. Considering this tendency, the spin-dependent hybridization with the ligand band would hardly occur in an FMS with a wide-band gap semiconductor. On the other hand, it has been pointed out theoretically that narrow-gap III-V semiconductors with strong covalency are appropriate for host compounds of both p-type and n-type FMSs [30, 31]. This scenario is consistent with the observations that n-type FMS (In,Fe)Sb with a smaller band gap shows higher $T_C$ than (In,Fe)As:Be [10] and p-type (Ba,K)(Zn,Mn)$_2$As$_2$ has higher $T_C$ than (Ga,Mn)As [32]. Since the IB states originate from strong *s-d* or *p-d* hybridization, the band structures proposed in the present study are likely to be applicable to other FMSs based on narrow gap semiconductors. Another important aspect here is that the band structure with the split-off IB changes with the carrier concentration. Then, additional carrier-doping is necessary to validate realization of carrier-induced ferromagnetism in FMSs aside from the TM dopants acting as donors or acceptors like Mn ions in (Ga,Mn)As. For example, Mn-doped FMSs with narrow-gap II-VI semiconductors co-doped with donor dopants will be candidates for n-type FMSs. Many FMS materials without carrier doping have been found to be paramagnetic so far. The emergence of carrier-induced ferromagnetism in such FMSs should be re-examined with additional carrier doping.

One of the most intriguing features here is that the energy position of the $\alpha$-IB depends on the CB filling, which means that the rigid-band picture is not applicable. The split-off $\alpha$-IB through the *s-d* hybridization in n-type (In,Fe)As discussed below is analogous to the split-off IB through the *p-d* hybridization in p-type (Ga,Mn)As [17], although the carrier type is different. Note that the strong hybridization is essential to form the split-off IB states in both the FMSs. As to the underlying physics, the formation of these IB states due to hybridization with ligands is also analogous to that of the Zhang-Rice singlet band [33, 34]. It is likely that the split-off IB can be formed in a spin-dependent way depending on the carrier concentration and strength of the *sp-d* hybridization. These arguments suggest that appearance of the hybridization-derived split-off IB states near the band edge of host semiconductor (VBM for *p*-type or CBM for *n*-type) is attributed to be a unified picture to realize the carrier-induced ferromagnetism in FMS materials irrespective to the carrier type.

**Conclusion**

In conclusion, we have conducted SX-ARPES measurements combined with SDFT calculations for the prototypical n-type FMS (In,Fe)As:Be in order to elucidate the nature of the n-type FMSs. The SX-ARPES study reveals the entire band structure including the Fe-3$d$ IBs and the host InAs ones, providing direct evidence for electron occupation of the InAs-derived CB. The Fe-3$d$ IB located near the CBM is formed through the hybridization with the ligand, which is the origin of the carrier-induced ferromagnetism in n-type (In,Fe)As. When the electron carriers occupy the CB, the partially filled IB splits off from the CBM leading to the large spin-polarization of the CB. The intriguing feature that its band structure is deformed depending on the band filling in the conduction band is unexpected from the ordinary rigid-band picture. The underlying physics of the formation of the split-off IB is found to be similar to that in p-type (Ga,Mn)As [17], suggesting a unified picture of ferromagnetism in FMSs irrespective to the carrier types. Our experimental findings suggest that the formation of the split-off IB state near valence-band maximum or conduction-band minimum is a key aspect to realize carrier-induced ferromagnetism in p-type or n-type FMSs, respectively.

**Acknowledgment**

This work was supported by Grants-in-Aid for Scientific Research (Nos. 15H02109, 17H04922, 18H05345) from JSPS, and CREST (JPMJCR1777) and PRESTO Program (JPMJPR19LB) of Japan Science and Technology Agency. J.M. and W.K. would like to thank the CEDAMNF (CZ.02.1.01/0.0/0.0/15_003/0000358) co-funded by the Ministry of Education, Youth and Sports of Czech Republic. This work was partially supported the Spintronics Research Network of Japan (Spin-RNJ). M.K. acknowledges support from

**References**
[1] H. Ohno, Science **281**, 951 (1998).
[2] S. A. Wolf, D. D. Awschalom, R. A. Buhrman, J. M. Daughton, S. von Molnár, M. L. Roukes, A. Y. Chtchelkanova, and D. M. Treger, Science **294**, 1488 (2001).
[3] I. Žutić, A. Matos-Abiague, B. Scharf, H. Dery, and K. Belashchenko, Mater. Today **22**, 85 (2019).
[4] T. Dietl, Nat. Mater. **9**, 965 (2010).
[5] M. Tanaka, S. Ohya, and P. N. Hai, Appl. Phys. Rev. **1**, 011102 (2014).
[6] T. Nakamura, L. D. Anh, Y. Hashimoto, S. Ohya, M. Tanaka, and S. Katsumoto, Phys. Rev. Lett. **122**, 107001 (2019).
[7] K. Takiguchi, L. D. Anh, T. Chiba, T. Koyama, D. Chiba and M. Tanaka, Nat. Phys. **15**, 1134 (2019).
[8] N. H. D. Khang, and P. N. Hai, J. Appl. Phys. **126**, 233903 (2019).
[9] N. T. Tu, P. N. Hai, L. D. Anh, and M. Tanaka, Appl. Phy.s Lett. **108**, 192401 (2016).
[10] A. V. Kudrin, Yu. A. Danilov1, V. P. Lesnikov, M. V. Dorokhin, O. V. Vikhrova, D. A. Pavlov, Yu. V. Usov, I. N. Antonov, R. N. Kriukov, A. V. Alaferdov, and N. A. Sobolev, J. Appl. Phys. **122**, 183901 (2017).
[11] L. D. Anh, P. N. Hai, Y. Kasahara, Y. Iwasa, and M. Tanaka, Phys. Rev. B **92**, 161201(R) (2015).
[12] L. D. Anh, P. N. Hai, and M. Tanaka, Nat. Comm. **7**, 13810 (2016).
[13] P. N. Hai, L. D. Anh, S. Mohan, T. Tamegai, M. Kodzuka, T. Ohkubo, K. Hono, and M. Tanaka, Appl. Phys. Lett. **101**, 182403 (2012).
[14] P. N. Hai, L. D. Anh, and M. Tanaka, Appl. Phys. Lett. **101**, 252410 (2012).
[15] N. D. Vu, T. Fukushima, K. Sato, and H. Katayama-Hoshida, Jpn. J. Appl. Phys. **53**, 110307 (2014).
[16] T. Dietl, H. Ohno, and F. Matsukura, Phys. Rev. B **63**, 195205 (2001).
[17] M. Kobayashi, I. Muneta, Y. Takeda, Y. Harada, A. Fujimori, J. Krempasky, T. Schmitt, S. Ohya, M. Tanaka, M. Oshima, and V. N. Strocov, Phys. Rev. B **89**, 205204 (2014).
[18] J. Krempasky, S. Muff, F. Bisti, M. Fanciulli, H. Volfová, A.P. Weber, N. Pilet, P. Warnicke, H. Ebert, J Braun, F. Bertran, V. V. Volobuev, J. Minár, G. Springholz, J. H. Dil, and V. N. Strocov, Nat. Comm. **7**, 13071 (2016).
[19] S. Sakamoto, Y. K. Wakabayashi, Y. Takeda, S.-i. Fujimori, H. Suzuki, Y. Ban, H. Yamagami, M. Tanaka, S. Ohya, and A. Fujimori, Phys. Rev. B **95**, 075203 (2017).
[20] V. N. Strocov, L. L. Lev, M. Kobayashi, C. Cancellieri, M.-A. Husanua, A. Chikina, N. B. M. Schröter, X. Wanga, J. A. Krieger, and Z. Salman, J. Electron Spectrosc. Relat. Phenom. **236**, 1 (2019).
[21] V. N. Strocov, X. Wang, M. Shi, M. Kobayashi, J. Krempasky, C. Hess, T. Schmitt and L. Patthey, J. Synchrotron Rad. **21**, 32 (2014).
[22] See Supplemental Material contains (I) Linear polarization dependence of the ARPES spectra, (II) Spectral line shape of Fe $L_3$ XAS, (III) Spin-density-functional-theory calculation, and (IV) Details of the calculation.

[23] G. K. Wertheim, J. E. Rowe, D. N. E. Buchanan, E. E. Chaban, A. F. Hebard, A. R. Kortan, A. V. Makhija, and R. C. Haddon, Science **252**, 1419 (1991).
[24] J. H. Weaver, P. J. Benning, F. Stepniak, and D. M. Poirier, J. Phys. Chem. Solids **53**, 1707 (1992).
[25] M. Kobayashi, L. D. Anh, P. N. Hai, Y. Takeda, S. Sakamoto, T. Kadono, T. Okane, Y. Saitoh, H. Yamagami, Y. Harada, M. Oshima, M. Tanaka, and A. Fujimori, Appl. Phys. Lett. **105**, 032403 (2014).
[26] P. W. Anderson, Phys. Rev. **124**, 41 (1964).
[27] J. R. Schrieffer, and P. A. Wolff, Phys. Rev. **149**, 491 (1966).
[28] J. Okabayashi, A. Kimura, O. Rader, T. Mizokawa, A. Fujimori, T. Hayashi, and M. Tanaka, Phys. Rev. B **58**, R4211 (1998).
[29] P. Manca, J. Phys. Chem. Solids **20**, 268 (1961).
[30] B. Gu, and S. Maekawa, Phys. Rev. B **94**, 155202 (2016).
[31] K. Sato, L. Bergqvist, J. Kudrnovský, P. H. Dederichs, O. Eriksson, I. Turek, B. Sanyal, G. Bouzerar, H. Katayama-Yoshida, V. A. Dinh, T. Fukushima, H. Kizaki, and R. Zeller, Rev. Mod. Phys. **82**, 1633 (2010).
[32] K. Zhao, Z. Deng, X. C. Wang, W. Han, J. L. Zhu, X. Li, Q. Q. Liu, R. C. Yu, T. Goko, B. Frandsen, Lian Liu, Fanlong Ning, Y. J. Uemura, H. Dabkowska, G. M. Luke, H. Luetkens, E. Morenzoni, S. R. Dunsiger, A. Senyshyn, P. Böni and C. Q. Jin, Nat. Comm. **4**, 1442 (2013).
[33] F. C. Zhang, and T. M. Rice, Phys. Rev. B **37**, 3759 (1988).
[34] K. Tsutsui, T. Tohyama, and S. Maekawa, Phys. Rev. Lett. **83**, 3705 (1999).

**Figure legend**

FIG. 1. SX-ARPES spectra of ferromagnetic (In,Fe)As:Be thin films. (a) Fermi surface mapping in the $k_z$-$k_x$ plane. $k_x$ indicates the momentum along the Γ-K-X direction. The solid curves are ***k***-space cuts for fixed photon energies. (b) Enlarged plot near the Γ point [$k_z$ = 15 (2π/c)]. The plotted area corresponds to the red square in panel a. The top and right plots are momentum distribution curves (MDCs) along the $k_x$ and $k_z$ cuts across the Γ point (dashed lines), respectively. (c) APRES intensity along the Γ-K-X line obtained with *p* polarization of incident X rays at *hν* = 908 eV. LH and SO denote the light-hole and split-off bands of the InAs matrix, respectively. The inset is a blow up around the Γ point. (d) Energy distribution curves (EDCs) around the Γ point.

FIG. 2. Resonant ARPES spectra of an (In,Fe)As:Be thin film. (a) Fe $L_3$ XAS spectrum. The constant-initial-state (CIS) spectrum at binding energy $E_B = 0.25$ eV corresponding to the $\alpha$-IB is also plotted. The arrows denote the excitation energy for the rARPES measurements. (b) rARPES spectra at the Fe $L_3$ edge. The left and right panels are the off- and on-resonance spectra, respectively. $\alpha$- and $\beta$-IBs are Fe 3$d$-derived impurity bands. (c) EDCs of the rARPES spectra. (Top) The on- and off-resonance EDCs at $k_x = 0$. (Bottom) Differences of the EDCs corresponding to the Fe 3$d$ component at various $k_x$. The vertical bars indicate a dispersive peak.

FIG. 3. Comparison of band structures of (In,Fe)As with and without doping. (a) Band dispersions near $E_F$ around the $\Gamma$ point along the $\Gamma$-K-X line in (In,Fe)As:Be. The IB intensity obtained by resonant ARPES is superimposed in the positive $k_x$ region. The dispersions of the LH and HH band are obtained from the spectra taken with $p$- and $s$-polarized incident X rays, respectively. (Right) EDC at the $\Gamma$ point and Fe-3$d$ PDOS. The colored areas denote the energy ranges of the Fe-3$d$ IBs. (b) Band dispersion near $E_F$ in InAs:Be, with the EDC at the $\Gamma$ point on the right. **c**, Band dispersion near $E_F$ around the $\Gamma$ point in (In,Fe)As as well as (In,Fe)As:Be (Fig.(a)).

FIG. 4. Band diagram of (In,Fe)As. (a),(b) Schematic DOS for ferromagnetic (In,Fe)As:Be and paramagnetic (In,Fe)As, respectively. DOS including the Fe 3$d$-derived IBs embedded in the VB and CB of the host InAs. The expected unoccupied Fe 3$d_\downarrow$ states are also shown. (c),(d) Band diagram for (In,Fe)As:Be and (In,Fe)As, respectively. Hybridization between Fe 3$d$ $e_\downarrow$ and CB leads to the formation of $\alpha$-IB and hybridization between Fe 3$d$ $t_{2\downarrow}$ and VB to the formation of $\beta$-IB. Dash-dotted lines are the band

dispersion of the host InAs without the hybridization. Electron carrier occupation of CB pushes down the partially filled $\alpha$-IB to be located below CBM, leading to the spin-splitting $\Delta_{ex}$ of the CB.

**Figure**

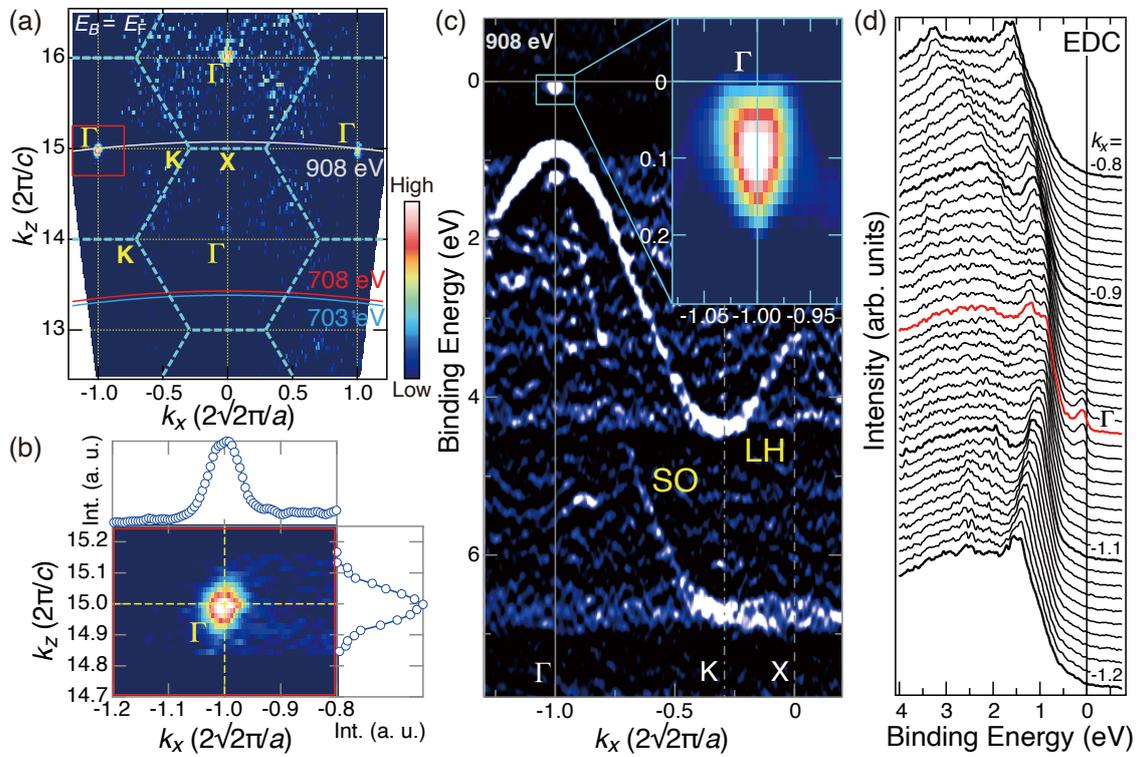

Figure 1

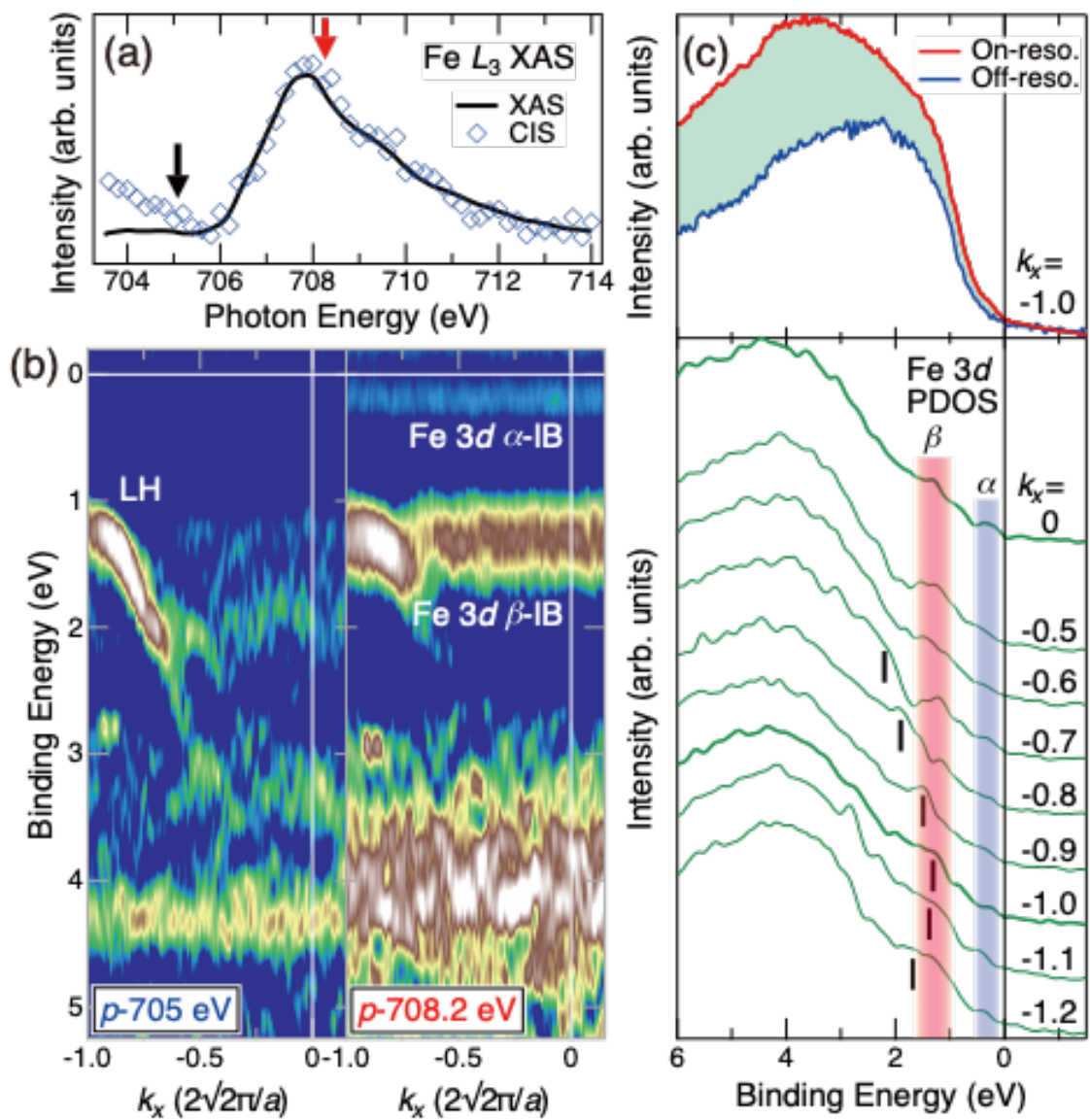

Figure 2

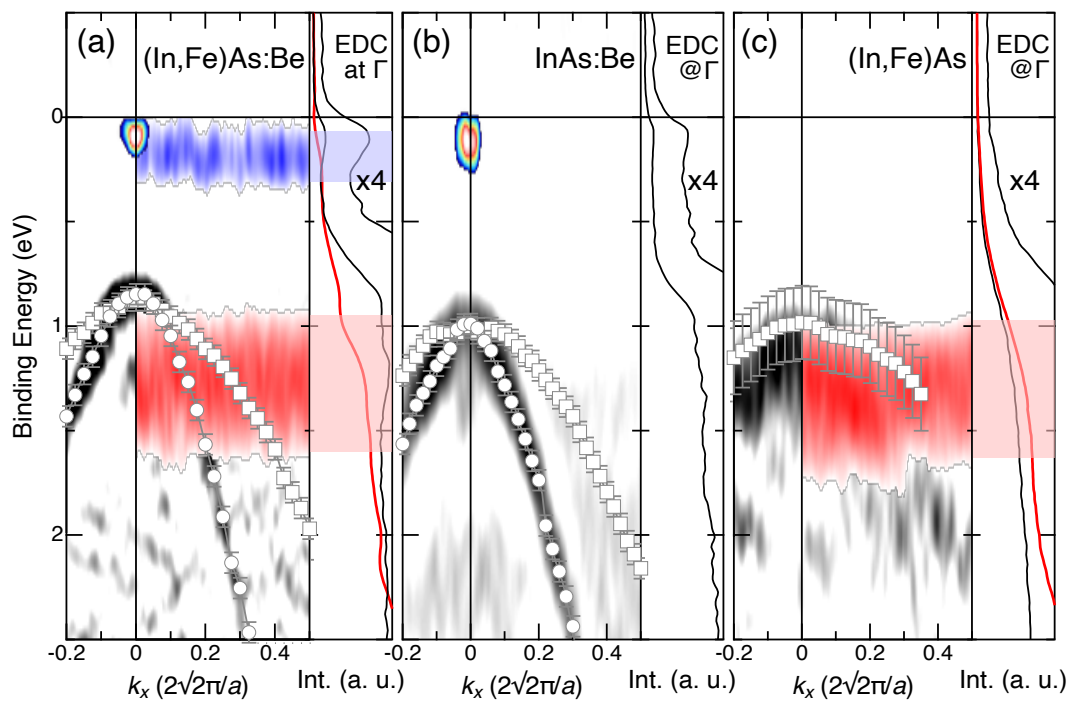

Figure 3

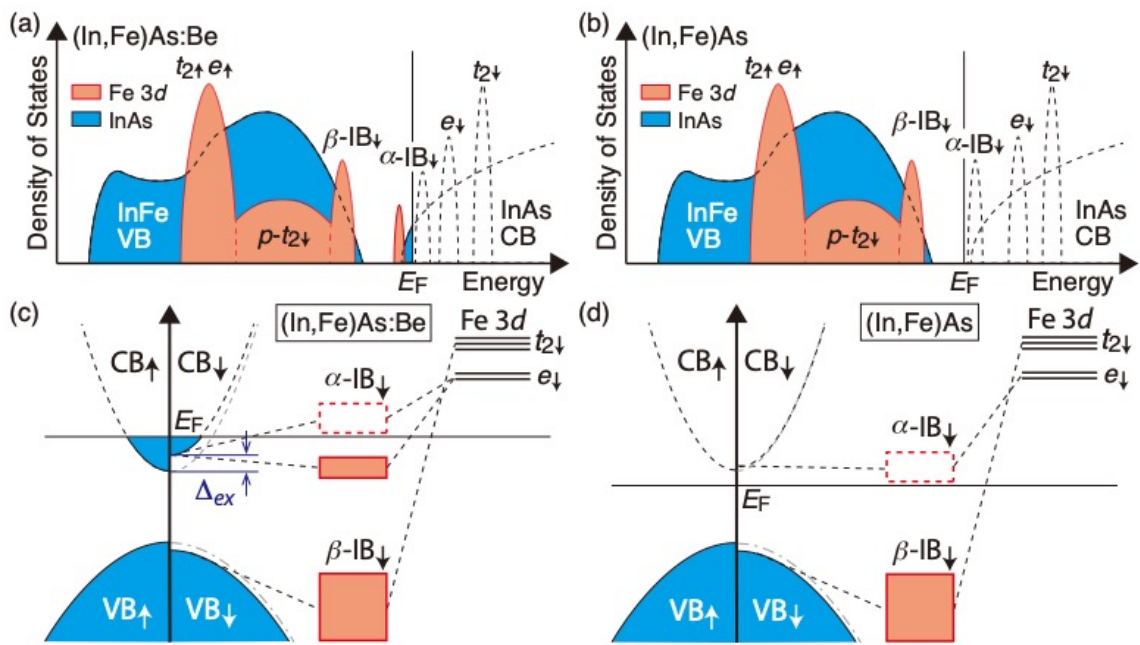

Figure 4

**Supplementary Material**

**I. Linear polarization dependence of the ARPES spectra**

The intensity of band dispersion depends on the polarization of the incident X rays reflecting the matrix element effects [1] and symmetry of the bands: The light-hole (LH) and split-off (SO) bands, and conduction-band minimum (CBM) of the host InAs can be observed with *p* polarization, and the heavy hole (HH) band with *s* polarization, as well as the polarization dependence of bands of GaAs [2, 3] (note, however, that the CBM of GaAs has not been observed in the previous study because the Fermi level ($E_F$) is located below the CBM in *p*-type GaAs). Figure S1 shows the angle-resolved photoemission spectroscopy (ARPES) spectrum along the Γ-K-X symmetry line taken with the *s* polarization. Only the HH band is clearly observed due to the matrix element effects. Since the linear polarization dependence of ARPES reflects the symmetry of the bands [1], the CBM observed only with *p* polarization has the same symmetry as the LH and SO bands, not as the HH band. Figure S2 shows Fe-$L_3$ resonant ARPES spectra taken with the $s$ polarization. Both the Fe 3$d$ $\alpha$- and $\beta$-impurity bands (IBs) can be observed in the on-resonance spectrum, while the intensities of the IBs are small and only the HH band can be observed in the off-resonance spectrum. Both the $\alpha$-IB and $\beta$-IB are active for the $p$ and $s$ polarizations, as shown in Figs. 2(b) and S2, indicating that the IBs possibly hybridize with the LH, HH, and SO bands of InAs.

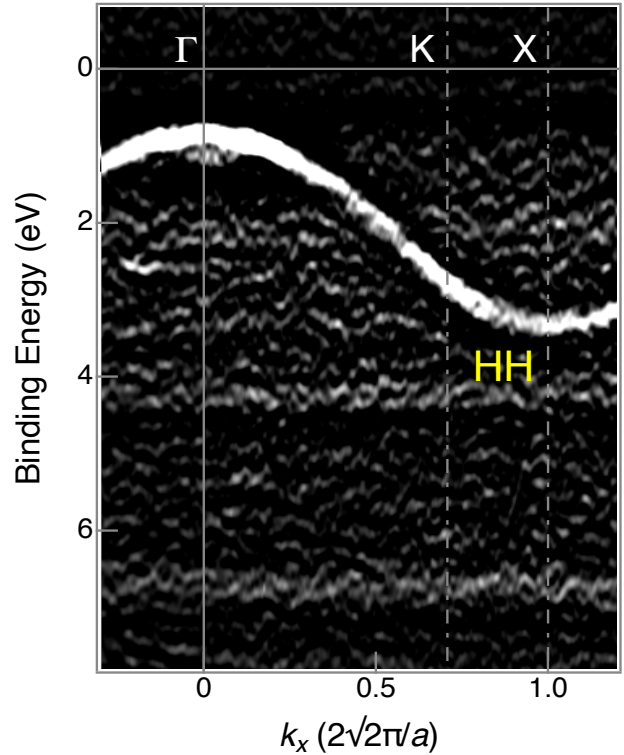

FIG. S1. APRES spectrum of (In,Fe)As obtained along the Γ-K-X line with *s* polarization. HH is the heavy-hole band of the host InAs. The incident photon energy for this measurement is of 908 eV as well as the ARPES spectrum taken with *p*-polarization shown in Fig. 2(c).

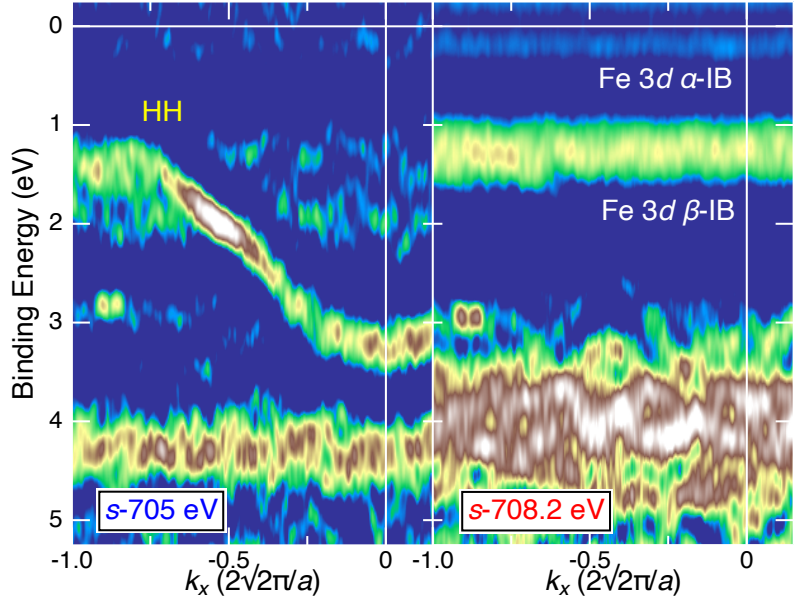

FIG. S2. Fe $L_3$ resonant ARPES spectra taken with $s$ polarization. The left and right panels are the off- and on-resonance spectra, respectively. IB is the Fe 3$d$-derived impurity band.

## II. Spectral line shape of Fe $L_3$ XAS

As shown in Fig. 2(e), the Fe $L_3$ x-ray absorption spectroscopy (XAS) spectrum shows a single peak around 708 eV with a small dip around $h\nu \sim$ 709.5 eV, different from multiple structures observed in Fe oxides [5], and the spectral line shape resembles those of Fe metal [6] and Fe pnictides [7]. The pre-edge structure around $h\nu \sim$ 705 eV comes from the In $M_2$ edge. Although the spectral feature is similar to that of Fe metal, our previous x-ray magnetic circular dichroism (XMCD) study confirms that the electronic structure of Fe in (In,Fe)As is different from Fe metal [8], excluding the possibility that the ferromagnetic property comes from aggregation of Fe metal clusters in (In,Fe)As. The negligibly weak signal of Fe oxide (at $h\nu \sim$ 709.5 eV) indicates that the amorphous As passivation layer well protected the (In,Fe)As:Be layer from oxidation.

## III. Spin-density-functional-theory calculations

To gain a basic insight into the band structure of (In,Fe)As:Be, we have performed spin-density-functional-theory calculation (SDFT) calculations for the band structures of (In,Fe)As. Figures S3(a) and S3(b) show the calculated band dispersions of (In,Fe)As for spin-up (majority) and spin-down (minority) states, respectively. Here, we focus on the spin-polarized states and their hybridization revealed by the SDFT calculation. The Fe $3d$ $e$ minority-spin ($e_↓$) band appears near the CBM in the calculation. Note that this $e_↓$ band shows a weak but finite dispersion, suggesting that the CBM states hybridize with the Fe $3d$ orbitals. Comparison with experiment as discussed below indicates that the observed $E_F$ is located just above the CBM (dash-dotted lines). Concomitantly, the down-spin states ($e_\downarrow$ and $t_{2\downarrow}$) should be shifted upwards to above $E_F$ due to the on-site Coulomb interaction at the Fe site, which is necessary to keep the Fe $d$-electron count within the realistic range but cannot be properly taken into account in SDFT. Also, an IB originating from the hybridization between the $e_\downarrow$ and CBM states appears below $E_F$. In this scenario, the spin direction of the Fe-$3d$ IB is opposite to the total magnetization, consistent with the observation of the spin-Esaki diode behavior using (In,Fe)As [9]. As shown in Fig. 1(b), the Fe $3d$ $t_2$ minority-spin states strongly hybridize with the valence band (VB) states of the InAs host below $E = -2$ eV (the $p$-$t_{2\downarrow}$ state). To see quantitatively how strongly the Fe $3d$ states are hybridized with the InAs band, the total density of states (DOS) and Fe $3d$ partial density-of-states (PDOS) are plotted in Fig. S3(c) and S3(d), respectively. The Fe $3d$ minority-spin states obviously contribute to the valence-band maximum (VBM) states, while the majority-spin band is largely located in the lower part of the VB (well below $E = -3.5$ eV) and has only negligible contributions to the VBM. Therefore, the SDFT calculation indicates that the minority-spin Fe $3d$ states strongly hybridize with the ligand VB and conduction band.

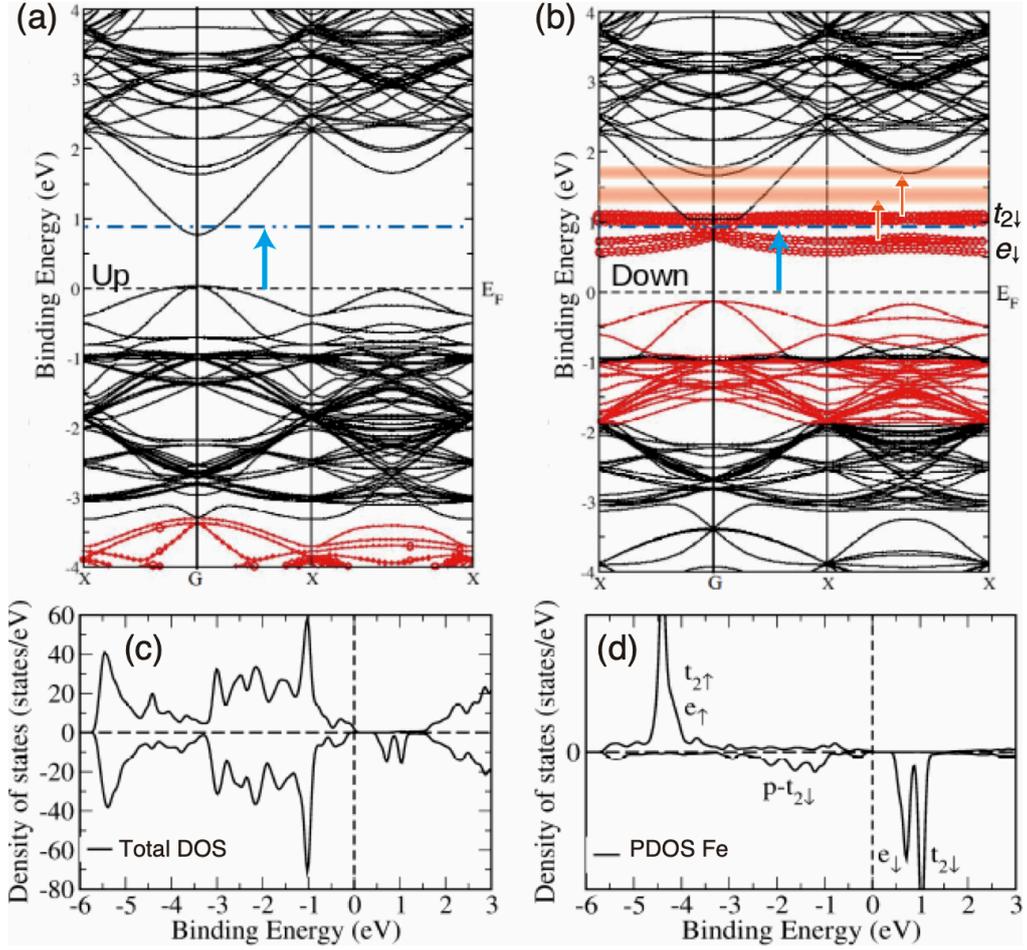

FIG. S3. Spin-density-functional-theory calculation for an $In_{31}Fe_1As_{32}$ supercell. (a),(b) Spin-resolved band structure. The weight of Fe $3d$ is indicated by the size of red circles. The blue dash-dotted line is the energy position of $E_F$ in the Be doped samples as observed experimentally. Note that as the $E_F$ is shifted upwards, the Fe $3d$ $t_{2↓}$ and $e_↓$ bands are also shifted upwards and remain located above $E_F$ due to Coulomb interaction as denoted by red arrows. The impurity band near $E_F$ ($\alpha$-IB) observed in the present work is considered to result from hybridization between the Fe $e_↓$ bands and the CBM. (c) DOS for the up-spin (upper curve) and down-spin states (lower curve). (d) Partial density of states for Fe $3d$.

**IV. Details of the calculation**

For the SDFT calculation of the electronic structure of $In_{0.9}Fe_{0.1}As$, we applied the all-electron augmented plane wave + local orbitals WIEN2K code. We employ the Engel-Vosko parameterization of a generalized gradient approximation (EV-GGA) [10]. This function has been successfully used in the past in studies of electronic band structure and density of states of InAs. Calculations for $In_{1-x}Fe_xAs$ has been performed within the supercell of 64 atoms, which nominally corresponds to $x = 0.3125$. The Brillouin zone was sampled by $5 \times 5 \times 5$ mesh. The wave functions in the atomic spheres were expanded up to an angular momentum of $l_{max} = 10$. The plane wave cutoff in the interstitial region was set so that $R_{MT} \times K_{MAX} = 7$ and the energy and charge convergence was set to $10^{-4}$ eV. Here, $R_{MT}$ and $K_{MAX}$ are the smallest atomic sphere radius and the largest K-vector of the plane wave expansion of the wave function, and the product $R_{MT} \times K_{MAX}$ describes the quality of the basis set used in the linearized augmented plane wave (LAPW) method.

**Reference**

[1] A. Damascelli, Z. Hussain, Z.-X. Shen, Rev. Mod. Phys. **75**, 473 (2003).

[2] M. Kobayashi, I. Muneta, T. Schmitt, L. Patthey, S. Ohya, M. Tanaka, M. Oshima, and V. N. Strocov, Appl. Phys. Lett. **101**, 242103 (2012).

[3] V. N. Strocov, X. Wang, M. Shi, M. Kobayashi, J. Krempasky, C. Hess, T. Schmitt and L. Patthey, J. Synchrotron Rad. **21**, 32 (2014).

[4] P. N. Hai, L. D. Anh, and M. Tanaka, Appl. Phys. Lett. **101**, 252410 (2012).

[5] T. Kataoka, M.Kobayashi, G. S. Song, Y. Sakamoto, A. Fujimori, F.-H. Chang, H.-J. Lin, D. J. Huang, C. T. Chen, S. K. Mandal, T. K. Nath, D. Karmakar, and I. Dasgupta, Jpn. J. Appl. Phys. **48**, 04C200 (2009).

[6] C. T. Chen, Y. U. Idzerda, H.-J. Lin, N. V. Smith, G. Meigs, E. Chaban, G. H. Ho, E. Pellegrin, and F. Sette, Phys. Rev. Lett. **75**, 152 (1995).

[7] W. L. Yang, A. P. Sorini, C-C. Chen, B. Moritz, W.-S. Lee, F. Vernay, P. Olalde-Velasco, J. D. Denlinger, B. Delley, J.-H. Chu, J. G. Analytis, I. R. Fisher, Z. A. Ren, J. Yang, W. Lu, Z. X. Zhao, J. van den Brink, Z. Hussain, Z.-X. Shen, and T. P. Devereaux, Phys. Rev. B **80**, 014508 (2009).

[8] M. Kobayashi, L. D. Anh, P. N. Hai, Y. Takeda, S. Sakamoto, T. Kadono, T. Okane, Y. Saitoh, H. Yamagami, Y. Harada, M. Oshima, M. Tanaka, and A. Fujimori, Appl. Phys. Lett. **105**, 032403 (2014).

[9] L. D. Anh, P. N. Hai, and M. Tanaka, Nat. Comm. **7**, 13810 (2016).

[10] E. Engel and S. H. Vosko, Phys. Rev. B **50**, 10498 (1994).